\documentclass[sigconf]{acmart}

\copyrightyear{2025}
\acmYear{2025}
\setcopyright{acmlicensed}\acmConference[CIKM '25]{Proceedings of the 34th ACM International Conference on Information and Knowledge Management}{November 10--14, 2025}{Seoul, Republic of Korea}
\acmBooktitle{Proceedings of the 34th ACM International Conference on Information and Knowledge Management (CIKM '25), November 10--14, 2025, Seoul, Republic of Korea}
\acmDOI{10.1145/3746252.3761579}
\acmISBN{979-8-4007-2040-6/2025/11}

\usepackage[mathscr]{euscript} 
\usepackage{multirow}
\usepackage{graphicx}    
\usepackage{subcaption}  
\usepackage{booktabs}
\usepackage{float}
\usepackage{xcolor} 
\usepackage{enumitem}
\usepackage{algorithm}
\usepackage{algpseudocode}
\usepackage{amsmath}
\setlist[itemize]{leftmargin=12pt, itemsep=1pt, topsep=1pt}

\newcommand{\PSRQ}{PSRQ}

\begin{document}

\title{Progressive Semantic Residual Quantization for Multimodal-Joint Interest Modeling in Music Recommendation}

\author{Shijia Wang}
\email{wangshijia1@corp.netease.com}
\affiliation{%
  \institution{NetEase Cloud Music}
  \city{Hangzhou}
  \country{China}
}

\author{Tianpei Ouyang}
\email{ouyangtianpei@corp.netease.com}
\affiliation{%
  \institution{NetEase Cloud Music \\Hangzhou Dianzi University}
  \city{Hangzhou}
  \country{China}
}

\author{Qiang Xiao}
\authornote{Corresponding author.}
\email{hzxiaoqiang@corp.netease.com}
\affiliation{%
  \institution{NetEase Cloud Music}
  \city{Hangzhou}
  \country{China}
}

\author{Dongjing~Wang}
\email{dongjing.wang@hdu.edu.cn}
\affiliation{%
  \institution{Hangzhou Dianzi University}
  \city{Hangzhou}
  \country{China}
}

\author{Yintao Ren}
\email{renyintao@corp.netease.com}
\affiliation{%
  \institution{NetEase Cloud Music}
  \city{Hangzhou}
  \country{China}
}

\author{Songpei Xu}
\email{xusongpei@corp.netease.com}
\affiliation{%
  \institution{NetEase Cloud Music}
  \city{Hangzhou}
  \country{China}
}

\author{Da Guo}
\email{guoda@corp.netease.com}
\affiliation{%
  \institution{NetEase Cloud Music}
  \city{Hangzhou}
  \country{China}
}

\author{Chuanjiang Luo}
\email{luochuanjiang03@corp.netease.com}
\affiliation{%
  \institution{NetEase Cloud Music}
  \city{Hangzhou}
  \country{China}
}


\renewcommand{\shortauthors}{Shijia Wang et al.}

\begin{abstract}
In music recommendation systems, multimodal interest learning is pivotal, which allows the model to capture nuanced preferences, including textual elements such as lyrics and various musical attributes such as different instruments and melodies. Recently, methods that incorporate multimodal content features through semantic IDs have achieved promising results. However, existing methods suffer from two critical limitations: 1) intra-modal semantic degradation, where residual-based quantization processes gradually decouple discrete IDs from original content semantics, leading to semantic drift;  and 2) inter-modal modeling gaps, where traditional fusion strategies either overlook modal-specific details or fail to capture cross-modal correlations, hindering comprehensive user interest modeling.
To address these challenges, we propose a novel multimodal recommendation framework with two stages. In the first stage, our Progressive Semantic Residual Quantization ({\PSRQ}) method generates modal-specific and modal-joint semantic IDs by explicitly preserving the prefix semantic feature. In the second stage, to model multimodal interest of users, a Multi-Codebook Cross-Attention (MCCA) network is designed to enable the model to simultaneously capture modal-specific interests and perceive cross-modal correlations. Extensive experiments on multiple real-world datasets demonstrate that our framework outperforms state-of-the-art baselines.
This framework has been deployed on one of China’s largest music streaming platforms, and online A/B tests confirm significant improvements in commercial metrics, underscoring its practical value for industrial-scale recommendation systems.
\end{abstract}

\ccsdesc[500]{Information systems~Recommender systems}

\keywords{Music Recommendation, Multimodal Representation, Residual Quantization, Semantic ID}


\maketitle

\section{Introduction}
In contemporary music streaming platforms, users exhibit varying preferences across different musical modalities, such as lyrics, instrumental, and melodic. Even among different demographic groups, the emphasis on modal interests can differ significantly. For instance, Fiore et al. \cite{10.1093/jmt/thw005} found that adults focus more on lyrics, while children prioritize melody. Further research \cite{unknown, 9402806} has indicated that different musical modalities can have distinct impacts on users' emotions. However, traditional recommendation models primarily rely on collaborative filtering \cite{1167344}, focus on modeling users' behavior preference \cite{din}, and lack the ability to learn multimodal interests.

\begin{figure}[t]
    \centering
    \includegraphics[width=\linewidth]{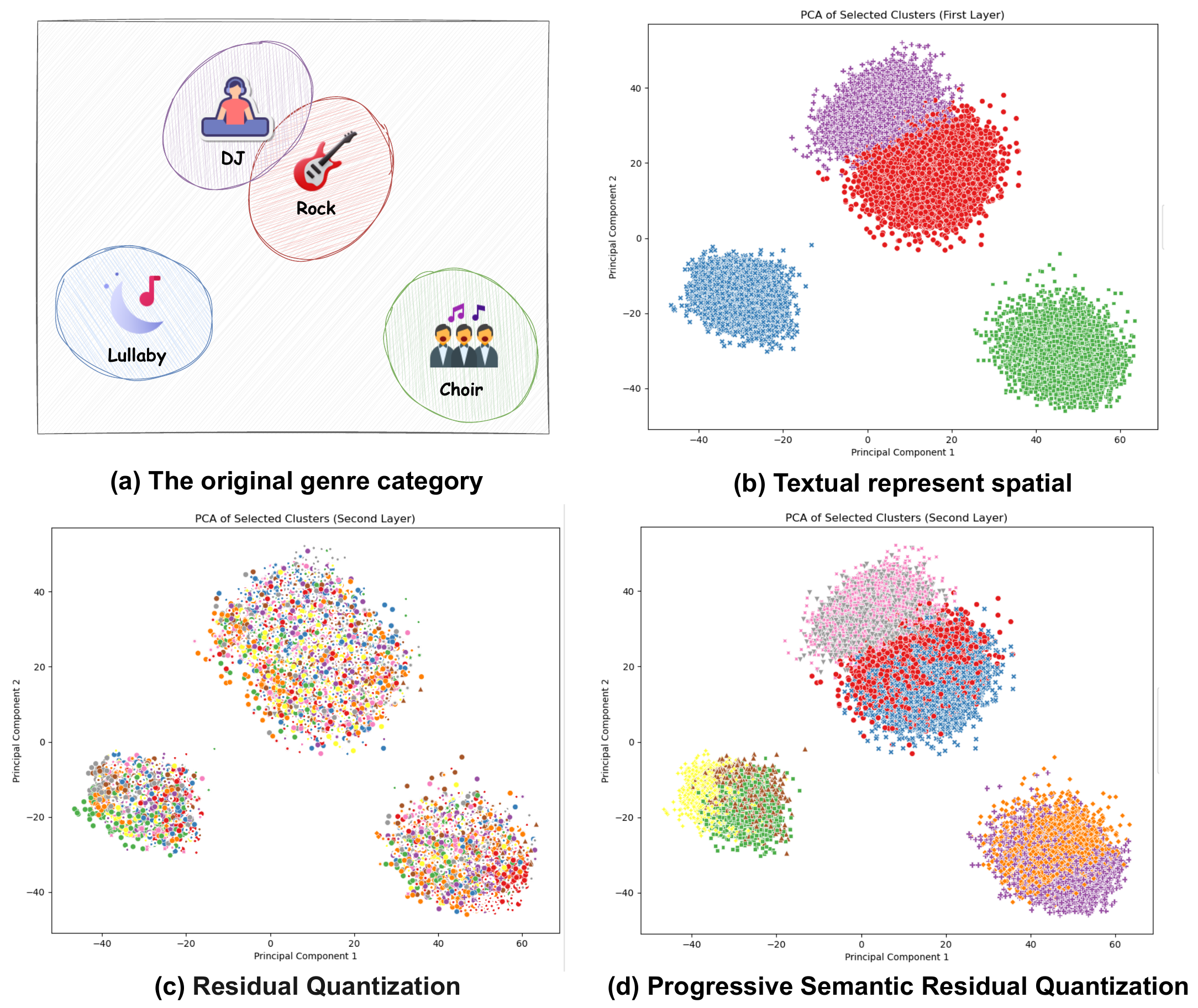}
    \Description{}
    \caption{The subfigure 1(b) illustrates the original text feature space of songs in subfigure 1(a), visualized by PCA. Subfigure 1(c) displays the clustering results of the second layer in traditional residual quantization. Subfigure 1(d) presents the clustering results of the second layer using the proposed PSRQ method.}
    \label{fig:instruction}
\end{figure}

In recent years, with the continuous advancement of multimodal feature extraction techniques, an increasing number of studies have applied multimodal information to fields such as short video and music recommendations~\cite{wang2025enhanced, VBPR, MMGCN, ModalityawareContrastive, xu2025efficientlargerecommendationmodel, 10884222}. These studies have shown that integrating multimodal information can significantly enhance the performance of recommendation systems by capturing the nuanced preferences of users. 
As researchers have delved deeper into multimodal recommendation, a central insight has emerged that semantic representation spaces of different modalities, including visual, textual, and acoustic features, exhibit pronounced disparities\cite{ImageBind, Alignbeforefuse, lin2023vila}. These differences can hinder the integration of cross-modal cues, which are crucial for understanding the subtleties of user preferences\cite{AlignRec, 10.1145/3583780.3614775, RAFAILIDIS201711}. In addition, traditional content representations are inherently static, as they are precomputed and fixed during training, posing a challenge as they cannot be optimized within an end-to-end recommendation framework. This limitation may impede the model's adaptability to complex interaction patterns and could result in slower convergence during training\cite{sheng2024Taobao}. The question of how to bridge multimodal representations with recommendation systems to enable end-to-end training has become a difficult problem.

Recently, quantization techniques have been widely applied in various fields and achieved remarkable results~\cite{Babenko2014Additive,li2021trq, Li_2017_ICCV}. Among them, VQ-Rec\cite{Hou2023VqRec} identified key challenges in vector quantization methods for recommender systems (VQ4Rec) and presented promising opportunities that can inspire future research in this emerging field. TIGER~\cite{Rajput2023tiger} further introduced the Residual Quantized Variational AutoEncoder (RQ-VAE~\cite{Lee_2022_CVPR}), officially opening the door of transformation of content feature into semantic IDs in the field of recommendation, through the application of the codebook~\cite{RQ2014, Lee_2022_CVPR}. Subsequently, numerous studies ~\cite{Singh2024Better, Li2025semantic} have demonstrated that semantic ID-based representation can bridge the aforementioned representation gap while endowing models with end-to-end adaptability for multimodal information. Furthermore, once the multimodal content of a cold-start item is mapped to a semantic ID, the item immediately inherits the learnable embedding of this ID from the codebook. This yields reliable and training-efficient representations, which significantly mitigate the challenges of data sparsity and cold-start problems \cite{QARM, M3CRS}.

Despite these advancements, two critical challenges remain:
\begin{itemize}
\item \textbf{intra-semantic of multimodal}: 
Current approaches rely purely on geometric similarity (e.g., Euclidean distance or cosine similarity between residuals) for quantization. While Residual Quantization (RQ)\cite{Lee_2022_CVPR} and RQ-VAE improve the accuracy of semantic IDs through iterative residual approximation, their layer-wise quantization process inherently decouples residual vectors from original semantic meanings, overlooks hierarchical semantic alignment—the deeper the quantization layer, the weaker the connection to the original semantics. As shown in Figure \ref{fig:instruction} (c), the residual vectors can lead to more diverse and discrete clustering results, but tend to overlook the associations with the original semantics. Consequently, the generated cluster IDs may deviate from the intended item semantics, leading to suboptimal recommendation performance. 

\item \textbf{inter-semantic of multimodal}: 
Existing paradigms such as QARM\cite{QARM} and OneRec\cite{deng2025onerecunifyingretrieverank} first fuse multimodal features via contrastive learning before quantization, which inevitably suppresses modal-unique signals critical for fine-grained user preference modeling. While M3CRS~\cite{M3CRS} preserves modal-specific characteristics through an independent embedding table, its isolated modeling the modal-specific interest of user fails to capture cross-modal synergies (e.g., audio-lyrics complementarity in music). However, in the context of recommendation systems, both aspects are of paramount importance\cite{sheng2024Taobao, AlignRec, M3CRS,chen2022breaking, zhou2025mdemodalitydiscriminationenhancement}. Therefore, the second challenge is how to simultaneously capture fine-grained modal preferences and exploit complementary cross-modal correlations for multimodal interest modeling based on semantic IDs.
\end{itemize}

To address these challenges, we propose a multimodal quantization-based recommendation framework that enhances both semantic fidelity and cross-modal interaction. In the feature engineering stage, we use a novel Progressive Semantic Residual Quantization ({\PSRQ}) method preprocesses multimodal embeddings by explicitly preserving prefix semantic feature, generating modal-specific and joint semantic IDs that maintain strong alignment with original semantics. Then, for the multimodal interest modeling of user, we introduce the Multi-Codebook Cross-Attention(MCCA) Network, which employs a shared modal-joint codebook as a cross-modal query to model multimodal embedding sequences. 
This approach operates end-to-end in the ranking stage\cite{10.1007/978-981-97-5575-2_26} of the recommendation system, jointly optimizing for semantic consistency and adaptive multimodal fusion to achieve superior recommendation performance.

%

In summary, the contributions of our study are as follows:
\begin{itemize}
\item We proposed a novel Progressive Semantic Residual Quantization method that constrains residual quantization with prefix semantics, enhancing semantic preserve.
\item We proposed a {Multi-Codebook Cross-Attention} Network for multimodal interest learning, simultaneously capturing modality specificity and cross-modal associations.
\item Extensive offline experiments on three real-world datasets and online A/B tests verified the effectiveness of proposed method, significantly improving cold start performance metrics.
\end{itemize}

\section{Related Work}
\subsection{Multimodal Representation for Recommendation}

In recent years, multimodal content features have achieved exciting results in enhancing recommendation systems. Early representative work, such as VBPR\cite{VBPR}, introduced visual features into the recommendation field using matrix factorization. MMGCN\cite{MMGCN} modeled the fine-grained modality of user preferences for each user-item bipartite graph of different modalities. To further improve multimodal based recommendation, the problem of multimodal interest representation fusion is crucial. A common approach is to fuse the multimodal embeddings of items through pre-training tasks. AlignRec\cite{AlignRec} pre-trained the visual-text alignment task using a mask-then-predict strategy. Sheng et al.~\cite{sheng2024Taobao} refined users' multimodal interest representations through SimTier for the multimodal representations after contrastive learning pre-training tasks, and used MAKE to address the issue of the difference in training epochs required for ID features versus multimodal representations.


\subsection{Quantitative Representation Learning for Recommendation}
Quantitative representation learning has recently attracted extensive attention from numerous scholars due to its ability to extract semantic information, and its effectiveness has been demonstrated in multiple fields. VQ-Rec\cite{Hou2023VqRec} converts text content vectors into sparse semantic ID representations through product quantization (PQ). TIGER\cite{Rajput2023tiger} further utilizes RQ-VAE to generate hierarchical semantic IDs based on the text content feature as item representations. Although RQ and RQ-VAE approximate the original embeddings through residuals to improve the accuracy of quantitative representations, they still face the hourglass problem\cite{kuai2024breaking}, resulting in uneven distribution in the discrete space and limited deep-level semantic associations. OneRec\cite{deng2025onerecunifyingretrieverank} enforces a uniform distribution of the number of elements in each layer of RQ through a multi-level balanced quantitative mechanism. Singh et al.~\cite{Singh2024Better} illustrated that Semantic IDs with Sentence Piece Models (SPM)\cite{kudo2018subword} are a more adaptive and efficient solution to represent item content and achieve better generalization outcomes. Furthermore, Zheng et al.~\cite{zheng2025enhancing} propose a prefix ngram parameterization method and prove that incorporating the hierarchical nature of clustering into the embedding table mapping is an effective measure.

\begin{figure}[t]
    \centering
    \includegraphics[width=\linewidth]{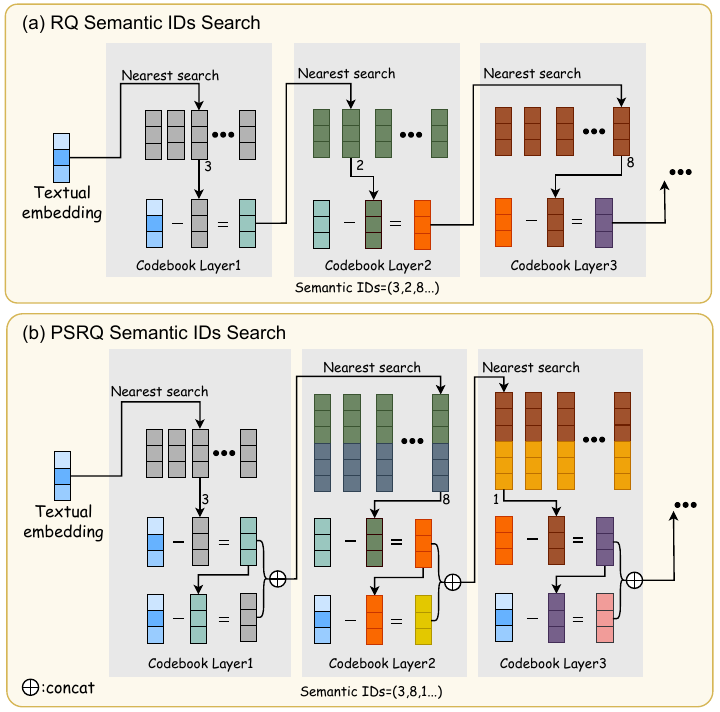}
    \Description{}
    \caption{Subfigure 2(a) and 2(b) provide a visual comparison of the semantic ID retrieval processes in the RQ and PSRQ codebooks.}
    \label{fig:PSRQ}
\end{figure}

\begin{figure*}[t]
    \centering
    \includegraphics[width=\linewidth]{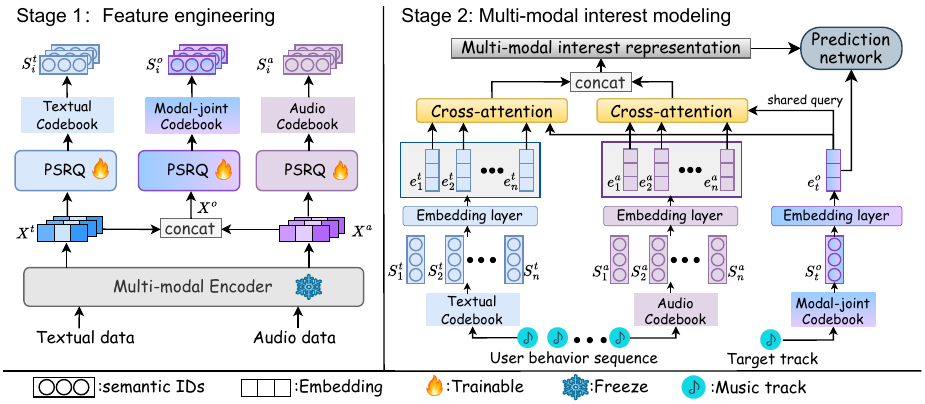}
    \Description{}
    \caption{The overall workflow of our recommendation framework. Although Stage 2 highlights only the core cross-attention component, we additionally conduct sequence modeling on both the modal-joint and the ID embedding sequences.}
    \label{fig:framework}
\end{figure*}

\section{Preliminary}
\textbf{\emph{Problem Definition.}} 
Let $\mathcal{U}$ and $\mathcal{I}$ denote the sets of users and items, respectively. And $|\mathcal{U}|$ and $|\mathcal{I}|$ denote user and item number. For all items, we obtain their multimodal content embeddings $X^m \in \mathbb{R}^{|\mathcal{I}| \times d}$ based on existing content feature extraction methods. Here, $m \in \{v, t, a\}$, where $v$ represents visual, $t$ represents textual, and $a$ represents audio, respectively. The specific content feature extraction methods are detailed in Section~\ref{sec:experiments}.
For each user $u \in \mathcal{U}$, we construct his historical behavior sequences $\mathcal{H}_u = \{i_1^h, i_2^h, \ldots, i_n^h\}$ based on positive interactions such as click, comment, or collect. In this sequence, $i_n^h\in\mathcal{I}$ denotes the item associated with the $n$-th interaction.
Our recommendation task involves predicting the probability $\hat{y}_{u,t}$ that the user $u$ will positively interact with the target item $i_{t}\in\mathcal{I}$. 

\textbf{\emph{Residual Quantization.}} 
In the conventional K-means-based Residual Quantization (RQ) process, each layer takes the residual vector from the previous layer as input and applies the K-means algorithm to obtain the cluster centers, which form the codebook for the layer.
For each kind of multimodal embeddings $X^m_i$:
\begin{equation}
\begin{split}
C_1 &= \text{K-means}( X^m_1, k), X^m_1 = X^m \\
C_2 &= \text{K-means}(X^m_2, k), X^t_2 = X^m_1 - \text{NearestRep}(X^m_1,C_1) \\
&\;\;\vdots \\
C_l &= \text{K-means}(X^m_{l}, k), X^m_{l}=X^m_{l-1}-\text{NearestRep}(X^m_{l-1},C_{l-1})
\end{split}
\end{equation}
where $l$ is the number of quantization layers, $C_l \in \mathbb{R}^{k\times d}$ are the generated cluster center embeddings of layer $l$, \( k \) is the number of cluster centers of K-means, and the $\text{NearestRep}(\cdot)$ denote the nearest representation search in the cluster center embeddings. The semantic IDs retrieval process of RQ is shown in Figure \ref{fig:PSRQ} (a).

\section{Methodology}
In this section, we elaborate on the components of our framework and its overall deployment workflow, depicted in Figure \ref{fig:framework}, which consists of two stages, feature engineering and downstream recommendation model training. 

\subsection{Progressive Semantic Residual Quantization}
In the feature engineering phase, inspired by previous work \cite{M3CRS, QARM}, we do not directly utilize original static content multimodal content embeddings $X^m \in \mathbb{R}^{|\mathcal{I}| \times d}$. Instead, we employ our proposed Progressive Semantic Residual Quantization(\PSRQ) method to map these embeddings to semantic ID representations. Different with RQ, the {\PSRQ} introduces a critical modification that the residual vector is differentiated from the original content feature vector and then concatenated with it for enhancing the retention of the original semantic information, as detailed in the following equations:
\begin{equation}
\begin{split}
X^m_1 &= X^m \\
C_1 &= \text{K-means}(X^m_1, k) \\
X^m_2 &= X^m - \text{NearestRep}(X^m_1, C_1) \\
C_2 &= \text{K-means}\bigl(X^m_2 \oplus (X^m - X^m_2),\; k\bigr) \\
&\;\;\vdots \\
X^m_{l} &= X^m_{l-1} - \text{NearestRep}(X^m_{l-1}, C_{l-1}) \\
C_l &= \text{K-means}\bigl(X^m_{l} \oplus (X^m - X^m_l),\; k\bigr)
\end{split}
\end{equation}
where \(\oplus\) denote concat operation and $C^1\in \mathbb{R}^{k\times d}$, $C^2$ to $C^l$ all $\in \mathbb{R}^{k\times 2d}$, and $m \in \{v, t, a\}$. In our online system, we only used textual and audio modal embeddings, that $m=\{t, a\}$. Specifically, for the modal-joint information, we concatenated $X^m$ as modal-joint embeddings $X^o\in \mathbb{R}^{|\mathcal{I}| \times 2d}$ to perform {\PSRQ} and generate modal-joint semantic IDs $S^o_i$ for each item, where $o$ represents multimodal joint information. 

Then for each item $i$, we retrieve the nearest cluster ID $id^t \in (0,1,\dots,k-1)$ from each quantization layer as the semantic IDs $S^m_i =[id^m_1, id^m_2, \dots, id^m_l]$. The different of semantic IDs retrieve process between conventional RQ and {\PSRQ} is shown in Figure \ref{fig:PSRQ}. 

\subsection{Multi-Codebook Cross-Attention}
Following the quantization of modal-specific and joint content representations, we integrate semantic IDs into collaborative filtering-based recommendation models. This integration enables the modeling of user multimodal interests, enhancing the generalization capability.
\subsubsection{Hierarchical Embedding Layer}
To enable end-to-end optimization of content features beyond the constraints of static representations, we do not use the original cluster centroid embeddings as semantic IDs embedding. Instead, within the embedding layer of our model, we use randomly initialized embedding tables for both the modal-specific and joint semantic IDs generated through the quantization codebooks. Specifically, for the modal-specific semantic IDs $S^m_i$ and modal-joint semantic IDs $S^o_i$ of each item, we allocate randomly initialized embedding tables denoted as $E^z \in \mathbb{R}^{k \times d^{\prime}}$ for each quantization layer, where $d^{\prime}$ is the embedding size and $z=\{t,a,o\}$, which $t$ represents textual, $a$ represents audio, and $o$ represents modal-joint information. Then we can retrieve the semantic ID embeddings of each layer and aggregate them as the final semantic representation ${e^z_i} \in \mathbb{R}^{d^{\prime}}$ of each item in the user history sequence.
\begin{equation}
    {e^z_i} = \sum_{j=1}^l \text{one-hot}(id_{ij}^z)\times{E^z_{j}}, id_{ij}^z\in S^z_i
\end{equation}
where one-hot is a commonly used option, which encode the $id_{ij}$ into a one-hot vector. As this method, we form the modal-specific and joint semantic embedding sequences $\{e_1^z, e_2^z, \ldots, e_n^z\}$. 

For the target item $i_t$, we only use the modal-joint codebook $E^o$ to obtain its
modal-joint semantic embedding ${e^o_t} \in \mathbb{R}^{d^{\prime}}$, which aims to capture cross-modal associations.

Furthermore, we integrate collaborative signals from the recommendation system by leveraging the sequence of ID embeddings \( \{e_1^r, e_2^r, \ldots, e_n^r\}\) and the target item ID embedding \(e_t^r \in \mathbb{R}^{d^{\prime}}\). This approach ensures that the model effectively captures the collaborative patterns within the data.

\subsubsection{Cross-Attention Layer}
After obtaining the user modal-specific and joint embedding sequences, we use the cross-attention mechanism to model the user modal-specific interests while capturing cross-modal correlations. The modal-joint semantic IDs embedding of the target item ${e^o_t}$ is utilized as a shared query to calculate the attention scores over the modal-specific and joint embedding sequences of the user history behaviors. 
This design is capable of alleviating the issue of inconsistent representation spaces across modalities to a certain extent, while also supporting the capture of cross-modal correlations. The multimodal interest representations ${h^z_u} \in \mathbb{R}^{d^{\prime}}$ are then computed using attention-weighted aggregation:
\begin{equation}
\begin{split}
{h^z_u} &= \sum_{j=1}^n \text{Cross-Attention}(Q={e^o_t}, K={e^z_j}, V={e^z_j}) \\
    &= \sum_{j=1}^n \text{Attention}({e^z_j}\oplus {e^o_t}) \cdot {e^z_j}\\
\end{split}
\end{equation}
where $\text{Attention}(\cdot)$ is a feed-forward network with output as the activation weight. 

To capture collaborative patterns, we also apply the attention mechanism to the collaborative embedding sequence $\mathcal{H}_u^r$ and then obtain the collaborative interest representation of the user:
\begin{equation}
    {h_u^r} = \sum_{j=1}^n \text{Attention}({e_j^r}\oplus {e_t^r}) \cdot {e_j^r}
\end{equation}
where $h_u^r$ is collaborative interest representation of the user. 

\subsection{Model Prediction \& Optimization\label{sec:logloss}}
To derive the probability of positive interaction between the user and the target item, we concatenate the multimodal interest representations ${h^z_u}$, collaborative interest representation $h_u^r$ with the target item's collaborative embedding ${e_t^r}$ and modal-joint semantic IDs embedding ${e^o_t}$. This concatenated vector is then fed into a Multi-layer Perceptron(MLP) for the predicted logit $\hat{y}_{j}$. Since the positive interaction prediction is a binary classification task, we employ cross-entropy loss as the objective function for model training and optimization:
\begin{equation}
  \mathcal{L} = -\frac{1}{N} \sum_{j = 1}^{N} y_{j} \log\sigma(\hat{y}_{j})+(1 - y_{j})\log(1 - \sigma(\hat{y}_{j}))
\end{equation}
where $N$ is the total number of training instances and \( y_j \in \{0, 1\} \) is the label for each sample. 

\section{Experiments}
In this section, we conduct a variety of offline experiments and online tests to evaluate the proposed method. Specifically, our aim is to address the following research questions.
\begin{itemize}
\item \textbf{RQ1} How does the proposed quantization-based recommendation framework compare in performance to the general and SOTA multimodal recommendation methods?

\item \textbf{RQ2} How does the semantic IDs generated from {\PSRQ} method perform in recommendation tasks compared to other quantization methods?

\item \textbf{RQ3} Does the modal-specific and modal-joint codebooks in MCCA effectiveness

\item \textbf{RQ4} How does the overall performance of the proposed method (PSRQ+MCCA) fare in real-world online scenarios?
\end{itemize}
\subsection{Experimental Settings}
\begin{table}[ht]\centering
    \caption{Statistics of datasets}
    \label{tab:dataset_info}
    \begin{tabular}{*{10}{c}}
        \toprule
       Dataset & \#Users & \#Items & \#Interactions \\
        \midrule
        Amazon baby & 81,423 & 33,652 & 230,444 \\
        Industrial & 4,926,656 & 1,387,247 & 8,696,093 \\
        Music4all & 14,127 & 99,596 & 2,597,382\\
        \bottomrule
    \end{tabular}
\end{table}
\subsubsection{Datasets\label{sec:experiments}}
To evaluate the performance of the proposed method, we conduct experiments on an industrial dataset and two public datasets, including Amazon Baby\cite{hou2024bridginglanguageitemsretrieval}, and Music4all\cite{Santana2020Music4AllAN}. 
Detailed data statistics and multimodal information of each dataset are presented in Table \ref{tab:dataset_info}. For all datasets, we mark items with fewer than 30 interactions as cold-start items.
\begin{table*}
  \caption{Performance of all baselines on various datasets is evaluated using AUC (for all items and cold-start items), with model fit assessed via Logloss. The top model is in bold, the second underlined. "\%Improv." denotes percentage-based relative improvement over the best baseline. "All AUC" is calculated on all items, while "Cold AUC" is for cold-start items (fewer than 30 interactions).
  }
  \label{tab:baseline compere}
  \begin{tabular*}{0.88\linewidth}{c|c c c| c c c| c c c}
    \toprule
    \multirow{2}{*}{Methods} & \multicolumn{3}{c|}{Amazon Baby} &  \multicolumn{3}{c|}{Industrial} & \multicolumn{3}{c}{Music4all} \\
    \cmidrule(lr){2-4}\cmidrule(lr){5-7} \cmidrule(lr){8-10}
   &  All AUC & Cold AUC & Logloss & All AUC & Cold AUC & Logloss & All AUC & Cold AUC & Logloss \\
    \midrule
    VBPR & 0.6466 & 0.5377 & 2.6145 & 0.7407 & 0.7229 & 1.4869 & 0.6217 & 0.5174 & 5.7460 \\
    SimTier+MAKE & 0.6213 & 0.5286 & 2.7974 & 0.7537 & \underline{0.7446} & 1.2503 & 0.6871 & 0.6041 & 3.6962 \\
    DIN & 0.6492 & 0.5487 & 2.5766 & 0.7599 & 0.7382 & 1.2188 & 0.7260 & 0.6699 & 2.9386 \\
    QARM & \underline{0.6557} & \underline{0.5681} & \underline{2.5686} & \underline{0.7628} & 0.7429 & \underline{1.2068} & \underline{0.7347} & \underline{0.7336} & \underline{2.9181} \\
    PSRQ+MCCA & \textbf{0.6573} & \textbf{0.5781} & \textbf{2.5564} & \textbf{0.7636} & \textbf{0.7535} & \textbf{1.2006} & \textbf{0.7347} & \textbf{0.7373} & \textbf{2.9070} \\
    \midrule
    \%Improv. & +0.24\% & +1.76\% & -0.47\% & +1.04\% & +1.20\% & -0.51\% & +0.00\% & +0.50\% & -0.38\% \\
    \bottomrule
  \end{tabular*}
\end{table*}

\begin{table*}
  \caption{Augment only textual semantic IDs to the DIN model to fairly assess the performance of each quantization approach. The top performer is in bold, and the second is underlined.
  }
  \label{tab:offline_quantization_results}
  \begin{tabular*}{0.84\linewidth}{c|c c c| c c c| c c c}
    \toprule
    \multirow{2}{*}{Methods} & \multicolumn{3}{c|}{Amazon Baby} &  \multicolumn{3}{c|}{Industrial} & \multicolumn{3}{c}{Music4all} \\
    \cmidrule(lr){2-4}\cmidrule(lr){5-7} \cmidrule(lr){8-10}
   &  All AUC & Cold AUC & Logloss & All AUC & Cold AUC & Logloss & All AUC & Cold AUC & Logloss \\
    \midrule
    DIN & 0.6492 & 0.5487 & 2.5766 & 0.7599 & 0.7382 & 1.2188 & 0.7260 & 0.6699 & 2.9386 \\
    +PQ & 0.6534 & 0.5515 & 2.5754 & 0.7617 & 0.7387 & \underline{1.2050} & 0.7313 & 0.7321 & 2.9234 \\
    +VQ & 0.6520 & \underline{0.5569} & 2.5725 & 0.7623 & \underline{0.7411} & 1.2087 & 0.7328 & 0.7317& \textbf{2.9146} \\
    +RQ & 0.6531 & 0.5546 & \textbf{2.5661} & \underline{0.7628} & 0.7334 & 1.2077 & 0.7329 & \underline{0.7327} & 2.9204 \\
    +RQ-VAE & \underline{0.6535} & 0.5533 & 2.5711 & 0.7620 & 0.7400 & 1.2081 & \underline{0.7338} & 0.7322 & 2.9184 \\
    +PSRQ & \textbf{0.6540} & \textbf{0.5610} & \underline{2.5705} & \textbf{0.7630} & \textbf{0.7442} & \textbf{1.2003} & \textbf{0.7345} & \textbf{0.7331} &  \underline{2.9183}\\
    \bottomrule
  \end{tabular*}
\end{table*}

\begin{itemize}
\item \textbf{Amazon Baby}: For the baby benchmark in the Amazon review dataset, we define interactions with ratings of 4 or higher as positive samples and those with ratings below 4 as negative samples. User interaction sequences are constructed based on timestamps. The image features of the items are extracted using the preexisting CNN-based method provided by the dataset. Textual features are obtained by the LLaMA3.2-1B\cite{malinovskii2024pushinglimitslargelanguage} model from the text composed of the product title, description, brand, and categorical information. 
\item \textbf{Industrial}: This dataset is collected from our online music platform over a one-week period.
To construct the training samples, we treat users' song collection interactions as positive samples and songs that users played but did not collect as negative samples; user historical interaction sequences are further built based on these positive interactions. For textual feature extraction, we leverage the Baichuan2-7B model \cite{yang2023baichuan2openlargescale} to generate textual embeddings from multi-source textual information, including song titles, genre tags, and lyrics. For audio feature extraction, we use the MERT-v1-95M model \cite{li2023mert} to extract audio embeddings directly from MP3-format audio files.
\item \textbf{Music4all}: 
We define repeatedly played songs as positive samples and songs played only once as negative samples. Text features are extracted by LLaMA3.2 from song titles, lyrics, genre tags, etc. The audio features are also sampled and extracted by the MERT model. 
\end{itemize}


\subsubsection{Implementation Details}
We implement all methods in TensorFlow 2, while the epoch number is set to one, the training batch sizes in three datasets are \{64, 512, 512\} and the learning rates are \{0.0005, 0.0001, 0.0001\}. 
The dimension $d^{\prime}$ of both the multimodal semantic IDs embedding and ID embedding is set to 64. The maximum length of the user history sequence is truncated to 20. For all models, we adopt the Adam~\cite{kingma2014adam} optimizer. Multimodal Large Language Models\cite{wang2024comprehensivereviewmultimodallarge} (MLLMs), including Baichuan2-7B, MERT-v1-95M, and LlaMA3.2-1B, are all deployed on NVIDIA A100 GPUs. For the proposed {\PSRQ} and other quantization methods, we set the number of clusters $k$ in Amazon Baby, Industrial, and Music4all at \{64, 256, 128\}, and the number of layers $l$ of RQ, PQ, RQ-VAE, and PSRQ are \{3,4,3,3\}. Furthermore, to ensure the fairness of the training process, we maintain consistent parameter scales, learning rates, and batch sizes across all models at different datasets.
\subsubsection{Evaluation Metrics} To comprehensively evaluate the performance of models, we adopt the AUC (Area Under the Curve) \cite{hanley1982meaning} as the primary evaluation metric. We define "All AUC" as the AUC metric evaluated on all items, while "Cold AUC" refers to the AUC metric for cold-start items. In addition, under the same model parameter magnitude, we provide the Logloss metric, with the specific formula detailed in Section \ref{sec:logloss}. 

\begin{table*}
  \caption{Ablate the role of modality-specific and cross-modal shared queries within the MCCA framework.}
  \label{tab:ablationstudy}
  \begin{tabular*}{0.87\linewidth}{c|c c c| c c c| c c c}
    \toprule
    \multirow{2}{*}{Methods} & \multicolumn{3}{c|}{Amazon Baby} &  \multicolumn{3}{c|}{Industrial} & \multicolumn{3}{c}{Music4all} \\
    \cmidrule(lr){2-4}\cmidrule(lr){5-7} \cmidrule(lr){8-10}
   &  All AUC & Cold AUC & Logloss & All AUC & Cold AUC & Logloss & All AUC & Cold AUC & Logloss \\
    \midrule
    w/o MSC & 0.6544 & 0.5549 & \underline{2.5594} & 0.7627 & 0.7424 & \textbf{1.1991} & 0.7333 & 0.7281 & 2.9184 \\
    w/o MJC & \underline{0.6555} & \underline{0.5619} & 2.5598 & \underline{0.7631} & \underline{0.7467} & \underline{1.2003} & \textbf{0.7353} & \underline{0.7364} & \underline{2.9089} \\
    PSRQ+MCCA & \textbf{0.6573} & \textbf{0.5781} & \textbf{2.5564} & \textbf{0.7636} & \textbf{0.7535} & 1.2006 & \underline{0.7347} & \textbf{0.7373} & \textbf{2.9070}\\
    \bottomrule
  \end{tabular*}
\end{table*}

\subsection{Offline Performance Comparison}

\subsubsection{Comparison with the Recommendation Models Proposed for Industry Scenarios(\textbf{RQ1})} To validate the effectiveness of the proposed framework, we compare the performance of our model with various baselines, including the latest SOTA models. 
\begin{itemize}
\item \textbf{DIN\cite{din}}: This model utilizes the attention mechanism to dynamically capture users’ interests from their historical behaviors. 
\item \textbf{VBPR\cite{VBPR}}: This model integrates the multimodal embeddings and ID
embeddings of each item as its representation and uses the matrix factorization (MF) framework to reconstruct the historical interactions between users and items.
\item \textbf{SimTier+MAKE\cite{sheng2024Taobao}}: An industrial-grade recommendation framework that combines similarity tiering and multimodal knowledge embedding to handle large-scale heterogeneous data. 
\item \textbf{QARM\cite{QARM}}: A recent SOTA approach that integrates contrastive learning and quantization techniques to enhance recommendation efficiency while preserving model expressiveness. 
\end{itemize}
As shown in Table \ref{tab:baseline compere}, the proposed PSRQ+MCCA model achieves superior performance across most experimental results on the three datasets, particularly excelling in the recommendation of cold-start items. QARM, which integrates contrastive learning and quantization techniques to enhance semantic generalization, attains the second-best performance across multiple datasets. DIN demonstrates good fit due to the thorough end-to-end training of ID embeddings, despite the lack of multimodal information. In contrast, SimTier+MAKE only outperformed VBPR, likely due to insufficient training within a single epoch.

\subsubsection{Comparison of Quantization Method (\textbf{RQ2})}
To rigorously assess the enhancement of the proposed PSRQ method over alternative quantization techniques in terms of content feature generalization and its impact on recommendation performance, we maintain identical batch sizes and learning rates alongside the DIN~\cite{din} model, with only textual modal embedding. This approach helps mitigate the influence of extraneous factors, ensuring a focused evaluation of the quantization methods themselves.
The quantization methods for comparison are as follows:
\begin{itemize}
\item \textbf{PQ\cite{PQ}}: It is a widely used technique that compresses high-dimensional vectors into a lower-dimensional space by dividing the vector into subvectors and quantizing each subvector independently.
\item \textbf{VQ\cite{vqvae}}: This method involves compressing high-dimensional vectors into a lower-dimensional space using a codebook generated by K-means clustering. 
\item \textbf{RQ\cite{RQ}}: RQ is also based on K-means clustering but focuses on quantization residuals iteratively to achieve more accurate approximations of the original vectors. 
\item \textbf{RQ-VAE\cite{Lee_2022_CVPR}}: This approach extends the concept of RQ by integrating it with an autoencoder architecture
, which helps in reconstructing rich semantic information. 
\end{itemize}
Referring to Table \ref{tab:offline_quantization_results}, all quantization methods have enhanced the performance of the DIN model.
Among them, PQ, lack of global semantic information, resulted in the poorest performance. VQ, despite employing only a single-layer quantization, achieved considerable improvements for cold-start items. Both RQ and RQ-VAE achieved the second-best results across multiple datasets. The PSRQ method demonstrated the overall best performance, including for all items and cold-start items in three datasets.

\subsubsection{Ablation Study of MCCA (\textbf{RQ3})}
To validate the effectiveness of MCCA, we conduct ablation studies, aiming to isolate the impact of modal-specific and joint:
\begin{itemize}
\item \textbf{w/o modal-specific codebooks(w/o MSC)}: We only used a modal-joint codebook during attention modeling, eliminating the dedicated modal-specific interest representations. This ablation tests whether removing granular modal semantics weakens the ability of modal to capture fine-grained user preferences across different content modalities.
\item \textbf{w/o modal-joint codebooks(w/o MJC)}: To assess the significance of cross-modal correlation modeling, we conducted experiments by excluding the shared modal-joint codebook for user's sequence and the shared query. Instead, we utilized modal-specific semantic IDs embeddings to serve as queries for the respective modality's semantic embedding sequences. This variant investigates whether the absence of a shared query mechanism hinders the capacity to exploit inter-modal dependencies, and affecting recommendation performance.
\end{itemize}

The results of the ablation study, as shown in Table \ref{tab:ablationstudy}, indicate that the MCCA framework outperforms on most datasets and metrics, validating the effectiveness of extracting cross-modal information through modal-joint codebooks and modeling each modality independently. Although in the Music4all dataset, MCCA's performance for all items is slightly inferior to that of w/o MJC, the improvements for cold-start items demonstrate that cross-modal associations are beneficial for enhancing the model's generalization performance, particularly for cold-start items.

\subsection{Online A/B Tests (RQ4)}
We conducted an A/B test of our online ranking model for our music streaming platform in February 2025, delivering song recommendations to tens of millions of users daily. The baseline model was the industry-standard Deep Learning Recommendation Model(DLRM\cite{naumov2019deeplearningrecommendationmodel}).
The experiment group enhanced the baseline by incorporating user multimodal interest representations, derived from PSRQ generated semantic IDs and MCCA.
Our core metrics include user engagement with music tracks, specifically through behaviors such as $collect$ and $full\_ play$. The $collect$ behaviors represent users' actions of adding songs to their favorite playlists, while the $full\_play$ behaviors signify that users have played the songs completely.
During the trial period, the experiment group saw a 2.81\% increase in $collect$ and a 0.95\% increase in $full\_play$ compared to the control group. For new tracks released within the last 30 days, the probabilities of $collect$ and $full\_play$ increased by 5.98\% and 2.2\%, respectively. In addition, the listening hours of the new tracks increased by 3.05\%.

\section{Conclusion}
In this work, we introduced a novel multimodal recommendation framework to address the persistent challenges of semantic degradation and cross-modal modeling gaps in music recommendation systems. Our Progressive Semantic Residual Quantization (PSRQ) method effectively preserves original semantic during quantization, while the Multi-Codebook Cross-Attention (MCCA) mechanism enables simultaneous capture of fine-grained multimodal interests and cross-modal correlations. Extensive experiments on multiple datasets demonstrated significant improvements, validating state-of-the-art performance of our framework. The successful deployment on a leading music streaming platform underscores its practical value in real-world scenarios. This study advances the field by bridging semantic fidelity and multimodal synergy, offering a scalable solution for industrial recommendation systems.


\section{Acknowledgments}
This research was supported by the Natural Science Foundation of Zhejiang Province under Grant No.LZ25F020010.


\bibliographystyle{ACM-Reference-Format}
\balance


\appendix









\end{document}